\documentclass[useAMS,usecolumn,usenatbib,usegraphicx]{mn2e}

\title[Galaxies and HI gas at $z=5$]{A search for galaxies in and around an HI overdense region at $z=5$\thanks{Based on data collected at Subaru Telescope and in part obtained from data archive at Astronomical Data Analysis Center, which are operated by the National Astronomical Observatory of Japan.}}
\author[Y. Matsuda et al.]{
\parbox[t]{\textwidth}{\vspace{-1cm}
Y.\ Matsuda,$^{\! 1}$\thanks{E-mail: yuichi.matsuda@durham.ac.uk} J. Richard,$^{\! 1}$ Ian Smail,$^{\! 1}$ N. Kashikawa,$^{\! 2}$ K. Shimasaku,$^{\! 3}$ B. L. Frye,$^{\! 4}$ T. Yamada,$^{\! 5}$ Y. Nakamura,$^{\! 5}$ T. Hayashino,$^{\! 6}$ T. Fujii $^{\! 6}$}\\\\
$^{1}$ Department of Physics, Durham University, South Road, Durham, DH1 3LE, UK\\
$^{2}$ National Astronomical Observatory of Japan, Mitaka, Tokyo 181-8588, Japan\\
$^{3}$ Department of Astronomy, School of Science, The University of Tokyo, Tokyo 113-0033, Japan\\
$^{4}$ Department of Physics and Astronomy, 2130 Fulton St., University of San Francisco, San Francisco, CA  94117, USA\\
$^{5}$ Astronomical Institute, Graduate School of Science, Tohoku University, Aramaki, Aoba-ku, Sendai 980-8578, Japan\\
$^{6}$ Research Center for Neutrino Science, Graduate School of Science, Tohoku Univsersity, Sendai 980-8578, Japan}
\begin{document}

\date{Accepted ... ; Received ... ; in original form ...}

\pagerange{\pageref{firstpage}--\pageref{lastpage}} \pubyear{2009}

\maketitle

\label{firstpage}

\begin{abstract}

We present the discovery of a large-scale structure of emission-line galaxies at redshift $z=4.86$ behind a massive cluster of galaxies, A1689. Previous spectroscopic observations of a galaxy, A1689-7.1 at $z=4.87$, near this structure, revealed a possible overdense region of inter-galactic medium (IGM) around the galaxy, which extends at least $\sim 80$ comoving Mpc along the line of sight. In order to investigate whether this $z\sim 5$ IGM overdense region contains a galaxy overdensity, we undertook narrow- and broad-band imaging observations around A1689-7.1 with Subaru/Suprime-Cam. We detected 51 candidate Ly$\alpha$ emitters (LAEs) at redshift $z=4.86\pm0.03$ in the $32 \times 24$ arcmin$^2$ field of view. After correction for lensing by the foreground cluster, we found a large-scale ($\sim 20 \times 60$ comoving Mpc) overdense region of galaxies around A1689-7.1 in the source plane at $z=4.86$. The densest peak in this region has an overdensity of $\delta \sim 4$, suggesting that this structure is probably a good candidate for a protocluster which may evolve into a massive cluster of galaxies in the present-day Universe. A1689-7.1 is located at the edge of this region, where the local galaxy density is $\sim 1.6$ times the mean density and is close to the density contrast in the IGM along the line of sight to A1689-7.1 estimated from the optical depth. The overdensities of galaxies we have found may suggest that at least some parts of the IGM overdense region have already started to form galaxies and moreover they relate to the formation of a protocluster. Although we lack information on the three dimensional distributions of both IGM and galaxy overdense regions, the similarity of the scales of both regions may suggest that the two are parts of a single large-scale structure, which would be an large edge-on sheet along the line of sight with a size of $\sim 20\times 60\times 80$ comoving Mpc.

\end{abstract}


\section{Introduction}

Galaxies are believed to interact frequently with the surrounding intergalactic medium (IGM) \citep[e.g.,][]{2005MNRAS.363....2K, 2007MNRAS.376.1465K}. There are several observational studies of these interactions using correlation between galaxies and IGM absorption-lines at $z\la 3$ \citep[e.g.,][]{2003ApJ...584...45A, 2005ApJ...629..636A, 2009arXiv0911.0368C}. While the studies of IGM absorption-lines usually need a bright background source, such as a QSO or a gamma-ray burst \citep[e.g., ][]{1998ARA&A..36..267R, 2005ApJ...634L..25C}, \citet{2008ApJ...685L...5F} claimed the existence of an overdense region of the IGM at $z\sim 5$ using a galaxy, A1689-7.1 at $z=4.866$, which is lensed by the foreground massive galaxy cluster A1689 at $z=0.187$. 

In the spectrum of A1689-7.1, the HI flux transmission at $6920 - 7100$ \AA\ (Ly$\alpha$ HI absorption-lines at $z = 4.69 - 4.84$) is only $\sim 30 \%$ of the average value at the same redshift derived from $z>5$ QSO spectra, suggesting that this overdensity of inter-galactic H I gas clouds around A1689-7.1 extends at least $\sim$80 comoving Mpc along the line of sight. Although individual H I absorption-lines could not be resolved at the resolution of their spectrum, they claimed that this structure is likely to be made by overlapping of many intergalactic H I gas clouds rather than a few overlapping of very high H I column density clouds, because of absence of metal absorption-lines from these clouds on the continuum long-ward of the Ly$\alpha$ emission. 

Such a large-scale IGM overdense region would provide us with an unique opportunity to study the formation of galaxies within baryonic large-scale structure at $z\sim 5$. Although there have been several high-$z$ emission-line galaxy searches targeting QSO absorption-line systems, the targets for most of these searches were absorption-line systems by very high H I column density clouds, such as damped Ly$\alpha$ absorbers rather than large-scale overdense regions of low H I column density clouds \citep[e.g.,][]{2001A&A...374..443F, 2001ApJ...554.1001F, 2004ApJ...602..545P}. \citet{2008ApJ...685L...5F} claimed that this IGM structure may be an example of the earliest phase of baryonic large-scale structure that subsequently evolve into the filamentary cosmic web of galaxies (including clusters of galaxies) observed in the present-day Universe. How can we test the possibility? Does the IGM overdense region already contain galaxies and even galaxy overdensities? If the overdensity of galaxies exists, what is the size, density, and morphology of the structure? In order to answer these questions, we undertook a wide-field, deep Ly$\alpha$ imaging survey for emission-line galaxies at $z\sim 4.8$ in the A1689 field. 

In this letter, we use AB magnitudes and adopt cosmological parameters, $\Omega_{\rm M} = 0.3$, $\Omega_{\Lambda} = 0.7$ and $H_0 = 70$ km s$^{-1}$ Mpc$^{-1}$. In this cosmology, the Universe at $z=4.86$ is 1.2 Gyr old and $1.0$ arcmin corresponds to a comoving length of 2.3 Mpc at $z=4.86$ \citep{2006PASP..118.1711W}.

\begin{figure}
\centering
  \includegraphics[scale=0.45]{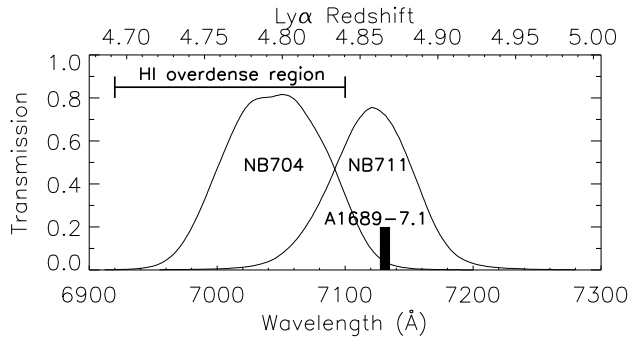}
  \caption{Transmission curves of $NB704$ and $NB711$ filters. The horizontal bar and vertical thick bar indicate the Ly$\alpha$ wavelength of the HI overdense region and the redshift of the lensed galaxy, A1689-7.1, $z=4.866$, respectively \citep{2008ApJ...685L...5F}.}
\end{figure}

\section[]{Observations and Data Reduction}

\begin{table}
\centering
\begin{minipage}{85mm}
  \caption{Summary of Observations }
  \begin{tabular}{@{}ccccc@{}}
  \hline
 Filter & $\lambda_{\rm cent}$/$\Delta \lambda$$^a$ & Exposure Time & $5 \sigma$ (lim)$^b$ & FWHM\\
  & (\AA/\AA) & (sec) & (AB mag) & (arcsec)\\
\hline
 $NB711$ & 7126/73 & 11700 ($900 \times 13$) & 25.7 & 0.8 \\
 $NB704^c$ & 7046/100 & 1800 ($900 \times 2$) & --- & --- \\
 $V$ & 5447/935 & 1920  ($480 \times 4$) & 26.3 & 0.8 \\
 $R$ & 6498/1124 & 5400  ($300 \times 18$) & 26.8 & 0.6 \\
 $i'$ & 7633/1489 & 7200  ($300 \times 24$) & 26.3 & 0.7 \\
 $Ri$ & 7066/2541 & -- & 26.6 & 0.8 \\
\hline
\end{tabular}
$^a$The central wavelength and FWHM of the filters.\\
$^b$The $5 \sigma$ limiting magnitudes within $1.6$ arcsec diameter apertures.\\
$^c$Significant ghosts appeared in $NB704$ images.\\
\end{minipage}
\end{table}

We obtained narrow- and broad-band imaging centred at ($\alpha$,$\delta$) = 13:11:29.5, $-$01:20:17 (J2000.0) on 2009 May 23 -- 24 (UT) using Suprime-Cam \citep{2002PASJ...54..833M} on the 8.2-m Subaru Telescope \citep{2004PASJ...56..381I} as an open use normal programme (S09A-104). We also used archival $V$-band data of the same field taken on 2001 Apr 24 (UT). Details of the observations are listed in Table~1. Our original plan was to use two narrow-band filters, $NB704$ and $NB711$  \citep[][see Fig.~1]{2003ApJ...582...60O, 2003ApJ...586L.111S, 2004ApJ...605L..93S}. We prioritized $NB704$ imaging because $NB704$ covers the redshift range for Ly$\alpha$ at $z\sim 4.76-4.84$ (46 comoving Mpc), which perfectly matches the redshift range of the IGM overdense region at $z \sim 4.69 - 4.84$. Unfortunately, the filter holder of $NB704$ had problems and significant ghosts appeared in our scientific images.\footnote{The filter holder problems were fixed by the staff of the Subaru Telescope immediately after the run.} We immediately changed filter from $NB704$ to $NB711$. Although the redshift coverage of $NB711$ ($z \sim 4.83-4.89$, 33 comoving Mpc) has a smaller overlap with the redshift range of the IGM overdense region than $NB704$, $NB711$ imaging is still useful to search for emission-line galaxies around A1689-7.1.

The raw data were reduced with {\sc sdfred} \citep{2002AJ....123...66Y, 2004ApJ...611..660O} and {\sc iraf}. We flat fielded using the median sky image after masking objects. We did background sky subtraction adopting the mesh size parameter of 64 pixels (13 arcsec) before combining the images. We calibrated the astrometry of the images using the 2MASS All-Sky Catalog of Point Sources \citep{2003tmc..book.....C}. Photometric calibration was obtained from the photometric and spectrophotometric standard stars, SA104, SA110-232, and Hz44 \citep{1990AJ.....99.1621O, 1992AJ....104..340L,2002AJ....123.2121S}. The magnitudes were corrected for Galactic extinction of $E(B-V)=0.03$ mag \citep{1998ApJ...500..525S}. The variation of the extinction in this field is small (peak to peak, $\pm 0.006$ mag) and thus it does not affect our results.

The combined images were aligned and smoothed with Gaussian kernels to match their seeing to a FWHM of $0.8$ arcsec. We made a $Ri$ image [$Ri=(R+i')/2$] for the continuum with an effective central wavelength similar to $NB711$. The total size of the field analyzed here is $32.1 \times 23.8$ arcmin$^2$ after removal of low S/N regions near the edges of the images. We also masked out the halos of the bright stars. The resultant total effective area is 723 arcmin$^2$. 

 Object detection and photometry were performed using {\sc SExtractor} version 2.5.0 \citep{1996A&AS..117..393B}. The object detections were made on the $NB711$ image using a Gaussian detection kernel with FWHM of $0.8$ arcsec. We detected 173,864 objects that had 5 connected pixels above $1.5 \sigma$ of the sky background. The magnitudes and colours are measured in $1.6$ arcsec diameter apertures. The magnitude limits of the observations are reported in Table~1.

\begin{figure}
\centering
  \includegraphics[scale=0.75]{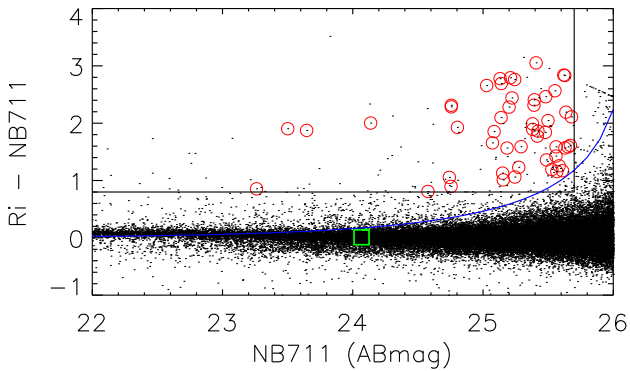}
  \caption{Colour-magnitude plot of $NB711$ vs $Ri-NB711$ for $NB711$ detected sample. The solid line indicates colour and magnitude limits ($Ri-NB711=0.8$ \& $NB711=25.7$). The red circles indicate 51 candidate LAEs. The green square indicates A1689-7.1. The blue curve indicates the $3 \sigma$ uncertainty of $Ri-NB711$ colour based on photometric errors of both $Ri$ and $NB711$ for objects with constant $f_{\nu}$ spectra. The emission-line objects that are not circled are considered to be low-$z$ interlopers from their broad-band colours (see text). All magnitudes and colours are measured with 1.6 arcsec diameter apertures.}
\end{figure}

\section[]{Results}

Fig.~2 shows a colour-magnitude plot for the $NB711$ detected sample. We selected 51 candidates as Ly$\alpha$ emitters (LAEs) down to $NB711=25.7$ ($5 \sigma$ limit) with the following criteria; 
\begin{enumerate}
\renewcommand{\theenumi}{(\arabic{enumi})}
\item $Ri - NB711 > 0.8$ ($EW_{\rm obs}>85$ \AA ),
\item $R-i'>0.5$, 
\item $i'-NB711>0$,
\item $V-i'>1.7$ or $V>27.3$ ($2 \sigma$) ,
\item $\Sigma > 3$, 
\end{enumerate}
where the $\Sigma$ is the ratio between the $NB711$ excess and the uncertainty of $Ri-NB711$ colour based on photometric errors of both $Ri$ and $NB711$ for objects with constant $f_{\nu}$ spectra. The criteria, (1), (2), and (3) are the same as used by \citet{2003ApJ...582...60O} and \citet{2003ApJ...586L.111S} to select $z=4.86$ LAEs with the same filter sets. The criterion (4) is applied to exclude the contamination from low-$z$ emitters, such as [OII]$\lambda 3727$ emitters at $z=0.91$, [OIII]$\lambda 5007$ emitters at $z=0.42$, and H$\alpha$ emitters at $z=0.09$. The $V-i'$ colour limit in (4) was determined using the expected colour of $z=4.9$ galaxies by \citet{2004AJ....127.2598S}. The criterion (5) was applied to exclude the contamination from non-emitters due to photometric errors. The properties of the 51 LAEs are listed in Table~A1 and displayed in Fig.~3. We note that A1689-7.1 was not selected as a LAE, because it failed to satisfy the criteria, (1), (3), and (5).

We corrected the measured magnitudes of LAEs to intrinsic (unlensed) magnitudes using the mass model of A1689 from \citet{2007ApJ...668..643L} which combines both strong- and weak-lensing constraints \citep[see also][]{2009arXiv0911.3302R}. While the magnification factor of A1689-7.1 equates to 2.5 magnitudes \citep{2008ApJ...685L...5F}, for our LAE sample, individual magnification factors range from 0 to 0.9 magnitudes. We found that the intrinsic $NB711$ magnitudes ($NB711_{int}$) of six LAEs are fainter than the $NB711$ magnitude limit ($NB711=25.7$). Lensing also reduces our surveyed area by the magnification factor, and we estimated a surface of $665$ arcmin$^2$ in the source plane, $8\%$ smaller than the observed field of view. We also estimated from the mass model the angular displacement of each candidate back to the source plane at $z=4.86$, typically 10 to 60 arcsec towards the centre of the cluster (the displacements from the measured positions, see Fig.~4) and perform all further calculations in the source plane.

We tested whether the entire field of view of the A1689 image (in the source plane) exhibits an overdensity by comparing to the $NB711$ surveys in Subaru Deep Field (SDF) \citep{2003ApJ...582...60O, 2003ApJ...586L.111S}. In the SDF, they detected 43 LAEs down to $NB711=25.5$ in $25 \times 45$ arcmin$^2$. The number density of LAEs in the SDF is $0.04\pm0.01$ arcmin$^{-2}$. The uncertainty is a $1\sigma$ Poisson error. Applying the same (intrinsic) magnitude limit to the A1689 field data, we detected 33 LAEs in $665$ arcmin$^2$. The number density of LAEs in the A1689 field is $0.05\pm0.01$ arcmin$^{-2}$. Thus the number density of LAEs in the A1689 field is similar to that in the SDF at least for a sub-sample of slightly bright LAEs (L(Ly$\alpha$)$> 1.7 \times 10^{42}$ erg s$^{-1}$). Thus the A1689 field does not show evidence for overdensity in the entire field of view if we use the SDF as a blank field.

\begin{figure*}
\centering
  \includegraphics[scale=0.93]{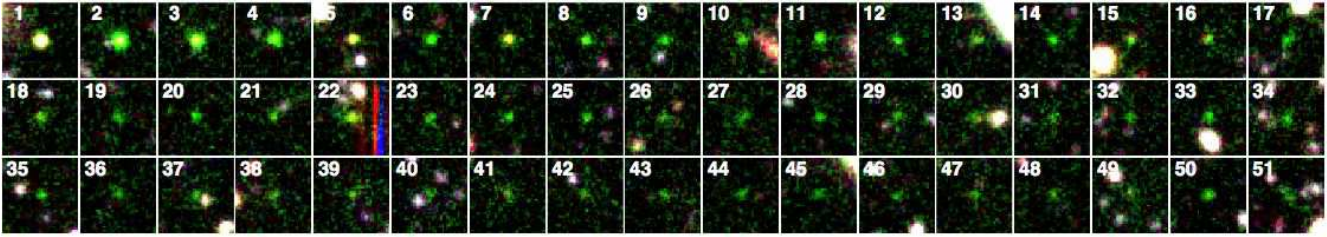}
  \caption{True colour images ($R$ for blue, $NB711$ for green, $i'$ for red) of LAEs. The size of each image is $8 \times 8$ arcsec$^2$ ($51 \times 51$ kpc$^2$ at $z=4.86$ in absence of lensing). The LAEs are ranked in terms of intrinsic (unlensed) $NB711$ magnitudes.}
\end{figure*}

\begin{figure*}
\centering
  \includegraphics[scale=.93]{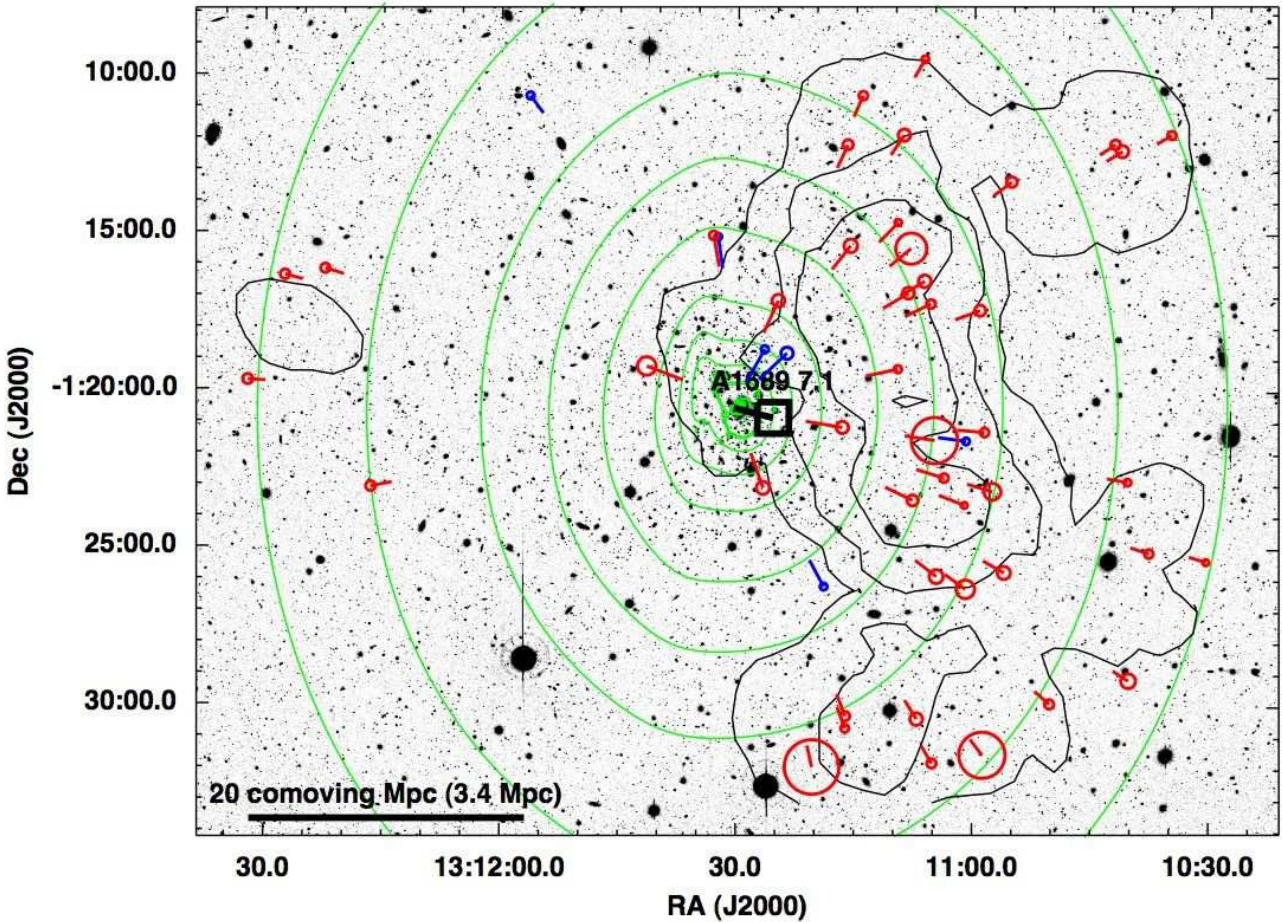}
  \caption{The sky distribution and smoothed density map (in the source plane) of candidate LAEs at $z\sim 4.86$ in the A1689 field. The gray scale image is the $NB711$ image. The thick horizontal bar indicates the angular scale of 20 comoving Mpc (3.4 Mpc in physical scale) at $z=4.86$. The red and blue circles indicate candidate LAEs with intrinsic (unlensed) $NB711$ magnitudes, $NB711_{int}\le25.7$ and $NB711_{int}>25.7$, respectively. The sizes of the circles are proportional to measured emission-line fluxes. The black square indicates A1689-7.1. The lines from these points indicate the positions in the source plane at $z=4.86$. The surface density map of LAEs (in the source plane) is generated with a top-hat smoothing kernel with a radius of $8$ comoving Mpc. The black contours indicate deviations from the mean density of this field, $\delta\equiv(n-\bar{n})/\bar{n}=0$, 1, 2, and 3, equivalent to significance levels of 0, 1.1, 2.2, and 3.3 $\sigma$. The green contours indicate lensing magnifications of 0.01, 0.02, 0.05, 0.1, 0.2, 0.5, 1.0, 2.0, and 5.0 mag.}. 
\end{figure*}

Fig.~4 shows the spatial distribution of the 51 candidate LAEs in the A1689 field and the smoothed density map in the source plane. To avoid effects from the spatial distribution of the lensing magnification, we made the density map using intrinsic magnitude limited sample of 45 LAEs with $NB711_{int} \le 25.7$ rather than the full 51 LAE sample. We used a top-hat smoothing kernel of 8 comoving Mpc radius, which is the same as that used in \citet{2003ApJ...586L.111S}. We defined high-density regions as the regions with the overdensity $\delta\equiv(n-\bar{n})/\bar{n}>0$. There is a large-scale overdense region with an extent of $\sim 8 \times 25$ arcmin$^2$ ($\sim 20 \times 60$ comoving Mpc) surrounding A1689-7.1. While the densest peak in this overdense region has an overdensity of $\delta = 3.9$, with a significance level of $4.3~ \sigma$, A1689-7.1 is located at the edge of this region, where the overdensity is $\delta = 0.6$.

We tested the reliability of the structure in the following way: (a) We slightly changed the criteria for LAEs (i.e. $\pm 0.1$ mag), but the properties of the structure do not change significantly. We conclude that the structure is stable against small changes in our sample definitions. (b) We divided the field of view into eastern and western parts, and selected low-$z$ emitters from $NB711$ detected sources down to $NB711=25.7$, satisfying criteria (1), (3), and (5), but not satisfying (2) or (4). The numbers of low-$z$ emitters in the eastern and western parts are $42\pm6$ and $55\pm7$, respectively. The uncertainties are $1\sigma$ Poisson errors. These numbers agree to within $30 \%$, showing that there is no clear systematic effect to select $NB711$ excess objects. (c) We also selected red objects down to $i'=25.7$, satisfying criteria (2) and (4). The numbers of red objects in the eastern and western parts are $1403\pm37$ and $1097\pm33$, respectively. These numbers also agree to within $30\%$, showing that there is no clear systematic effect to select red objects with $R-i'$ and $V-i'$ colours. In contrast to low-$z$ emitters and red objects, the numbers of LAEs in the eastern and western parts are $8\pm3$ and $43\pm7$, respectivley. Thus systematic errors of colours in the field of view should not affect the large-scale structure of LAEs significantly.

\section[]{Discussion and Conclusions}

What does the large-scale structure of galaxies (i.e., LAEs) we have found evolve into? This large-scale structure of galaxies is a good candidate for a protocluster. Numerical simulations of structure formation suggested that most ($>80\%$) of structures with a size of $\sim 20-40$ comoving Mpc and with a peak overdensity of $\delta \ge3$ at $z\sim 5-6$ will evolve into rich clusters ($M\ge 10^{14} h^{-1} M_{\sun}$) in the present-day Universe \citep{2006ApJ...646L...5S, 2009MNRAS.394..577O}. This structure has a size of $\sim 20 \times 60$ comoving Mpc and the densest region has an overdensity of $\delta \sim 4$. Hence we conclude that this structure is probably a good candidate for a protocluster.

How does the galaxy overdensity relate to the IGM overdensity in this structure? If we assume the optical depth in the IGM is proportional to the square of local gas density over the mean gas density [$\tau \propto (\rho/\bar{\rho})^2$] \citep[e.g.,][]{1998ARA&A..36..267R}, then the gas density in the IGM overdense region would be estimated to be $\sim 1.3$ times the mean gas density at $z\sim 5$ by using the optical depth from \citet{2008ApJ...685L...5F}. The local galaxy density at the position of A1689-7.1 is $\sim 1.6$ times the mean density, and hence it is close to that of the IGM. However, our current observations may have only probed the back-part of the IGM overdense region, and thus it is possible that the galaxy density will be higher in the foreground along the line of sight to A1689-7.1. Although there is a small overlap of the observed redshift ranges of the IGM and galaxy structures ($\Delta z \sim 0.01$), the current results suggest that at least some parts of the IGM overdense region may have already started to form galaxies and moreover they are probably related to the formation of a protocluster.

Is it possible that these galaxy and IGM overdense regions are parts of a single large-scale structure? Unfortunately, at present, we lack information on the three dimensional distributions of both IGM and galaxy structures. Without spectroscopy, it is unclear whether the structure is concentrated in a particular redshift or smoothly extends to the whole redshift range ($z\sim 4.83-4.89$). Similarly, it is still unclear whether or not the IGM overdense region extends in a transverse direction or to higher redshift than that of A1689-7.1. Nevertheless, the similarity of the scales of both structures (a few times ten comoving Mpc) may suggest that the two are parts of a single structure. If the IGM and galaxy overdense regions are parts of a single structure, this would be an large edge-on sheet along the line of sight with a size of $\sim 20\times 60\times 80$ comoving Mpc. The future observations of LAEs at $z\sim 4.79$ with $NB704$ will be a direct test of this possibility.

\section*{Acknowledgments}

We thank the anonymous referee for insightful comments which have significantly improved the paper. We thank Tom Theuns, Mark Swinbank, Jim Geach, Ichi Tanaka, Fumiaki Nakata, and Tomoki Saito for help and useful discussions. YM and IRS acknowledge support from STFC. JR acknowledges support from an European Union Marie-Curie Fellowship. BF acknowledges support from the Science Foundation Ireland Research Frontiers Programme grant PHY008.

\appendix

\section{Source Catalog}

\begin{table*}
\centering
\begin{minipage}{162mm}
 \caption{Properties of the 51 candidate $z=4.86$ LAEs in A1689 field$^a$}
 \begin{tabular}{@{}ccrrcccccccr}
  \hline
ID$^c$ & Coordinate (J2000)$^c$ & \multicolumn{2}{c}{Offset$^d$} & $NB711$ & $Ri$ & $V$ & $R$ & $i'$ & $NB711_{int}$ & log L$_{\rm {Ly}\alpha}$ & $EW_{obs}$ \\
LAE & (h:m:s) (d:m:s) & \multicolumn{2}{c}{(arcsec)} & \multicolumn{6}{c}{(ABmag)} & (erg s$^{-1}$) & (\AA) \\
   \hline
1 & 13:10:58.70 $-$01:31:33.0 &  23.2 &  30.2 & 23.26 & 24.12 & 25.97 & 24.63 & 23.69 & 23.28 & 43.04 & 93\\
2 & 13:11:20.32 $-$01:31:56.9 &   9.4 &  40.8 & 23.50 & 25.40 & 26.88 & 25.80 & 25.03 & 23.53 & 43.11 & 415\\
3 & 13:11:04.92 $-$01:21:33.2 &  57.1 &   8.3 & 23.65 & 25.52 & 27.62 & 26.17 & 25.04 & 23.74 & 43.00 & 400\\
4 & 13:11:08.02 $-$01:15:27.5 &  41.1 & -34.2 & 24.14 & 26.14 & 28.00 & 26.93 & 25.65 & 24.21 & 42.84 & 472\\
5 & 13:10:56.04 $-$01:25:45.0 &  40.0 &  22.6 & 24.58 & 25.39 & 27.08 & 26.06 & 24.90 & 24.61 & 42.48 & 85\\
6 & 13:11:00.93 $-$01:26:16.7 &  38.8 &  28.0 & 24.76 & 27.07 & 28.00 & 27.89 & 26.53 & 24.80 & 42.62 & 707\\
7 & 13:11:08.33 $-$01:16:52.6 &  48.6 & -29.5 & 24.75 & 25.65 & 28.00 & 26.59 & 25.09 & 24.85 & 42.41 & 100\\
8 & 13:10:57.56 $-$01:23:10.5 &  48.0 &  14.2 & 24.80 & 26.73 & 27.83 & 27.50 & 26.23 & 24.85 & 42.58 & 428\\
9 & 13:10:40.21 $-$01:29:10.2 &  29.4 &  19.0 & 25.03 & 27.68 & 28.00 & 28.36 & 27.25 & 25.04 & 42.54 & 1137\\
10 & 13:10:41.23 $-$01:12:20.5 &  30.4 & -19.0 & 25.07 & 26.73 & 27.75 & 27.31 & 26.28 & 25.09 & 42.46 & 301\\
11 & 13:11:41.52 $-$01:19:14.1 & -69.0 & -25.9 & 24.76 & 27.04 & 28.00 & 28.07 & 26.51 & 25.10 & 42.51 & 682\\
12 & 13:11:07.07 $-$01:30:24.4 &  22.7 &  35.9 & 25.09 & 26.94 & 28.00 & 27.26 & 26.66 & 25.12 & 42.46 & 389\\
13 & 13:12:22.38 $-$01:16:09.8 & -35.5 & -11.0 & 25.15 & 26.28 & 27.96 & 26.73 & 25.90 & 25.16 & 42.34 & 144\\
14 & 13:11:04.63 $-$01:25:52.7 &  40.3 &  30.9 & 25.13 & 27.91 & 28.00 & 28.67 & 27.27 & 25.19 & 42.49 & 1357\\
15 & 13:11:03.72 $-$01:22:44.7 &  53.1 &  16.2 & 25.16 & 26.17 & 27.66 & 26.72 & 25.74 & 25.23 & 42.28 & 120\\
16 & 13:10:34.98 $-$01:11:49.3 &  28.6 & -17.0 & 25.24 & 26.31 & 27.61 & 26.85 & 25.88 & 25.25 & 42.30 & 130\\
17 & 13:11:15.69 $-$01:15:22.8 &  35.5 & -45.4 & 25.14 & 27.83 & 28.00 & 28.44 & 27.61 & 25.26 & 42.46 & 1197\\
18 & 13:11:07.70 $-$01:23:27.5 &  52.3 & 24.7 & 25.18 & 26.75 & 28.00 & 28.00 & 26.15 & 25.28 & 42.36 & 267\\
19 & 13:11:08.96 $-$01:11:51.4 &  25.3 & -37.9 & 25.24 & 28.00 & 28.00 & 28.52 & 27.65 & 25.29 & 42.45 & 1326\\
20 & 13:11:06.27 $-$01:16:29.7 &  46.5 & -28.6 & 25.21 & 28.00 & 27.92 & 28.67 & 27.45 & 25.29 & 42.45 & 1391\\
21 & 13:12:27.48 $-$01:16:21.8 & -33.9 &  -9.2 & 25.29 & 26.88 & 27.49 & 27.51 & 26.47 & 25.30 & 42.36 & 276\\
22 & 13:11:26.68 $-$01:23:05.3 &  23.8 &  66.6 & 24.74 & 25.80 & 28.00 & 26.76 & 25.25 & 25.36 & 42.26 & 129\\
23 & 13:12:32.17 $-$01:19:42.2 & -34.1 &  -1.7 & 25.38 & 27.37 & 28.00 & 28.67 & 26.70 & 25.39 & 42.36 & 463\\
24 & 13:12:16.51 $-$01:23:05.0 & -40.2 &   8.0 & 25.38 & 27.27 & 27.89 & 27.64 & 27.00 & 25.40 & 42.36 & 409\\
25 & 13:10:55.28 $-$01:13:19.8 &  33.9 & -25.9 & 25.39 & 27.80 & 28.00 & 28.67 & 27.10 & 25.42 & 42.38 & 803\\
26 & 13:11:05.11 $-$01:31:49.0 &  20.1 &  33.7 & 25.42 & 27.19 & 28.00 & 27.94 & 26.76 & 25.44 & 42.32 & 351\\
27 & 13:11:16.10 $-$01:12:10.4 &  20.3 & -44.1 & 25.39 & 27.71 & 28.00 & 28.16 & 27.31 & 25.45 & 42.36 & 708\\
28 & 13:10:59.21 $-$01:17:25.2 &  47.7 & -17.6 & 25.41 & 28.46 & 28.00 & 28.67 & 28.01 & 25.46 & 42.40 & 2132\\
29 & 13:11:16.13 $-$01:30:19.5 &  16.2 &  41.8 & 25.42 & 27.28 & 28.00 & 28.25 & 26.72 & 25.47 & 42.32 & 392\\
30 & 13:11:33.08 $-$01:15:04.2 & -10.3 & -59.8 & 25.27 & 26.50 & 28.00 & 27.26 & 26.01 & 25.47 & 42.26 & 168\\
31 & 13:10:37.64 $-$01:25:07.6 &  35.4 &  11.5 & 25.48 & 27.32 & 28.00 & 27.94 & 26.85 & 25.49 & 42.32 & 383\\
32 & 13:10:42.13 $-$01:12:07.7 &  30.1 & -19.7 & 25.48 & 27.94 & 28.00 & 28.67 & 27.36 & 25.49 & 42.36 & 865\\
33 & 13:11:24.84 $-$01:17:08.0 &  27.7 & -61.6 & 25.14 & 27.24 & 27.65 & 27.86 & 26.83 & 25.50 & 42.34 & 533\\
34 & 13:11:16.65 $-$01:21:09.3 &  69.7 &  11.5 & 25.22 & 27.66 & 28.00 & 28.67 & 27.08 & 25.53 & 42.34 & 840\\
35 & 13:11:14.19 $-$01:10:36.4 &  17.7 & -40.5 & 25.50 & 27.54 & 28.00 & 28.41 & 26.99 & 25.54 & 42.32 & 498\\
36 & 13:10:50.21 $-$01:29:55.2 &  28.4 &  24.8 & 25.55 & 28.12 & 28.00 & 28.67 & 27.26 & 25.57 & 42.32 & 999\\
37 & 13:11:06.34 $-$01:09:26.5 &  20.1 & -34.6 & 25.56 & 26.99 & 27.37 & 27.25 & 26.74 & 25.59 & 42.23 & 221\\
38 & 13:11:09.67 $-$01:14:38.1 &  36.2 & -37.9 & 25.53 & 26.72 & 28.00 & 27.25 & 26.32 & 25.60 & 42.18 & 159\\
39 & 13:10:30.41 $-$01:25:23.2 &  32.6  &  9.9 & 25.61 & 26.78 & 28.00 & 27.36 & 26.35 & 25.62 & 42.18 & 154\\
40 & 13:11:09.63 $-$01:19:17.7 &  60.8 & -13.3 & 25.49 & 26.85 & 28.00 & 27.12 & 26.54 & 25.62 & 42.20 & 202\\
41 & 13:11:01.11 $-$01:23:36.6 &  48.7 &  18.9 & 25.57 & 26.72 & 28.00 & 27.27 & 26.30 & 25.63 & 42.18 & 151\\
42 & 13:10:40.40 $-$01:22:51.6 &  39.0 &   6.8 & 25.63 & 27.20 & 28.00 & 27.88 & 26.77 & 25.65 & 42.23 & 268\\
43 & 13:11:16.15 $-$01:30:42.8 &  15.3  & 41.2 & 25.64 & 27.83 & 28.00 & 28.67 & 27.22 & 25.68 & 42.28 & 604\\
44 & 13:10:58.57 $-$01:21:17.7 &  51.9  &  4.6 & 25.63 & 28.46 & 27.66 & 28.67 & 28.01 & 25.68 & 42.30 & 1480\\
45 & 13:11:05.46 $-$01:17:12.8 &  49.7 & -24.3 & 25.62 & 28.46 & 28.00 & 28.67 & 28.01 & 25.70 & 42.28 & 1497\\
46 & 13:11:56.43 $-$01:10:39.4 & -24.6 & -33.3 & 25.68 & 27.79 & 28.00 & 28.37 & 27.36 & 25.71 & 42.26 & 542\\
47 & 13:11:00.93 $-$01:21:34.6 &  53.6 &   6.9 & 25.67 & 27.28 & 28.00 & 28.37 & 26.70 & 25.74 & 42.20 & 284\\
48 & 13:11:18.93 $-$01:26:12.7 &  27.2  & 50.3 & 25.65 & 27.24 & 28.00 & 28.13 & 26.74 & 25.78 & 42.18 & 274\\
49 & 13:11:32.51 $-$01:15:06.7 &  -8.5 & -60.3 & 25.58 & 26.84 & 28.00 & 27.33 & 26.45 & 25.79 & 42.11 & 174\\
50 & 13:11:23.72 $-$01:18:48.3 &  53.0  &-50.7 & 25.20 & 27.48 & 28.00 & 28.67 & 26.85 & 25.85 & 42.20 & 678\\
51 & 13:11:26.53 $-$01:18:40.9 &  35.4 & -63.3 & 25.56 & 27.15 & 27.55 & 27.57 & 26.85 & 26.42 & 41.92 & 275\\
\hline
 \end{tabular}
$^a$ The properties are based on photometry with $1.6$ arcsec diameter apertures. Magnitudes below $1 \sigma$ are replaced to $1 \sigma$ upper limits.\\
$^b$ The LAEs are ranked in terms of intrinsic (unlensed) $NB711$ magnitude, $NB711_{int}$.\\
$^c$ The measured positions of LAEs.\\
$^d$ Offsets ($\delta$RA and $\delta$Dec in arcsec) of LAEs on the source plane. The positive values show offsets to East and North directions. 
\end{minipage}
\end{table*}

\label{lastpage}

\end{document}